\documentclass[
reprint,
nofootinbib,
aps,
pra,
amsmath,
amssymb,
]{revtex4-2}

\usepackage{graphicx}
\usepackage{multirow}
\usepackage{placeins}
\usepackage{xcolor}

\usepackage{braket}
\usepackage{mathrsfs}

\graphicspath{{img/}}

\DeclareMathOperator{\Tr}{\operatorname{Tr}}
\newtheorem{hypothesis}{Hypothesis}

\usepackage[
  colorlinks=true,
  citecolor=blue
]{hyperref}

\begin{document}

\title{Analytic formulae for nonlocal nonstabiliserness in bipartite systems of qutrits and ququints}

\date{\today}

\author{Giorgio Busoni}
\email{giorgio.busoni@adelaide.edu.au}
\affiliation{
  University of Adelaide, ARC Centre of Excellence for Dark Matter Particle Physics \& CSSM, Department of Physics, Adelaide SA 5000 Australia
}

\author{John Gargalionis}
\email{john.gargalionis@adelaide.edu.au}
\affiliation{
  University of Adelaide, ARC Centre of Excellence for Dark Matter Particle Physics \& CSSM, Department of Physics, Adelaide SA 5000 Australia
}

\author{Ewan N.\ V.\ Wallace}
\email{ewan.n.wallace@adelaide.edu.au}
\affiliation{
  University of Adelaide, ARC Centre of Excellence for Dark Matter Particle Physics \& CSSM, Department of Physics, Adelaide SA 5000 Australia
}

\author{Martin J.\ White}
\email{martin.white@adelaide.edu.au}
\affiliation{
  University of Adelaide, ARC Centre of Excellence for Dark Matter Particle Physics \& CSSM, Department of Physics, Adelaide SA 5000 Australia
}

\begin{abstract}
  We conjecture analytic expressions for the non-local magic of bipartite pure qudit states of prime local dimension. Our construction relies on the Schmidt-aligned state attaining the minimum over local unitaries, a hypothesis that we support with numerical evidence for pairs of qutrits and ququints. For composite local dimensions, we find that the analogous expressions do not in general reproduce the global minimum, but can still provide computationally cheap approximations to the non-local magic. We also find that relations between non-local magic and entanglement diagnostics that hold for two qubits generally do not extend to qutrit and higher-dimensional systems. As a few-body application, we evaluate the qutrit non-local magic generated by elastic pion scattering, treating hadronic isospin as the qutrit. We find that $p$-wave charged pion scattering realises the maximum non-local magic allowed in the system just below the $\rho$-meson resonance.
\end{abstract}

\maketitle

\section{Introduction}

Quantum computation based on higher-dimensional systems, or \emph{qudits}, has recently gained increasing attention as a route to enhance gate efficiency, noise resilience, and information density~\cite{9069177}. Among these, qutrits (local dimension $N=3$) and ququints ($N=5$) have emerged as leading candidates for near-term implementations. In any quantum computational system, identifying and quantifying the non-stabiliser content of states is essential, since non-stabiliser, or \emph{magic}, states represent a fundamental resource enabling universal quantum computation beyond the Clifford group~\cite{Gottesman:1998hu,Gottesman:1997zz,Aaronson:2004xuh,Beverland:2019jej,Bravyi:2016yhm,Bravyi:2018ugg,Bravyi:2015mry,Bravyi:2004isx,Bu:2019qed,Campbell:2011jhn,Gu:2024ure,Gu:2024qvn,Gu:2023ylw,Nakagawa:2025wjd,Liu:2020yso,Seddon:2021rij,True:2022ypf,Yoganathan:2019ppg,Zhou:2020scv,Bejan:2023zqm,Koh:2015dzu,Bouland:2018fbp,Zhang:2024fyp}.

Several approaches have been developed to quantify magic, e.g.~\cite{Howard:2017maw,Beverland:2019jej,Liu:2020yso,Leone:2021rzd}. Recently, a basis-independent measure of non-stabiliserness, the non-local non-stabiliserness, has been introduced~\cite{Qian:2025oit,Cao:2024nrx,Cao:2023mzo,Cepollaro:2024sod}. It is defined through the minimisation of a magic monotone $M$ over all local unitary transformations acting independently on two subsystems $A$ and $B$:
\begin{equation}
\mathscr{M}_{AB}(| \Psi_{AB} \rangle) = \min_{U_A \otimes U_B} \left[ M(U_A \otimes U_B | \Psi_{AB} \rangle) \right] \ ,
\end{equation}
where we use $\mathscr{M}$ for the non-local non-stabiliserness, which we here refer to as the non-local magic (NLM). This quantity shares the same symmetries as entanglement measures, making it a resource of central importance for quantum computation. Indeed, for a system of two pure qubits it can be directly related to the flatness of the entanglement spectrum~\cite{Tirrito:2023fnw,Robin:2025ymq,Iannotti2025entanglement}, and so a non-zero value for $\mathscr{M}$ signals the presence of both magic and entanglement in the state. For bipartite pure qubit systems, the minimisation over local unitaries admits an analytic solution in terms of invariants of the $\mathrm{U}(2)\otimes \mathrm{U}(2)$ group, but an analogous formulation for qudits with local dimension $N>2$ has remained unavailable.

In this work, we generalise the analytic construction of the NLM to bipartite pure qudit systems of prime dimension. Our approach is to expand the density matrix on the basis of tensor-product generalised Pauli operators and identify quartic invariants of these coefficients under local unitary transformations. Our derivation assumes that the NLM minimum is attained by a Schmidt-aligned state, a hypothesis we call \textit{Schmidt attainment}. Subject to this hypothesis, we obtain closed analytic expressions for the NLM of two qutrits and two ququints. Direct optimisation over $U_A \otimes U_B$ provides strong numerical support for this identification at $N=3,5$, motivating the conjecture that Schmidt attainment extends to all prime local dimensions. In contrast, for the composite dimension $N=4$, the Schmidt-aligned ansatz is not globally optimal throughout parameter space, although it often provides a good and computationally cheap approximation. We also report the spectra of the reduced density matrices corresponding to minimal and maximal NLM values for $N=3,5$, providing a geometric interpretation of the non-stabiliserness landscape in these cases.

The results presented here offer a rapid, analytic and basis-independent means to evaluate non-stabiliserness in prime-dimensional bipartite pure systems of low dimension, relevant to current experimental and theoretical efforts in qudit-based quantum computing, and recent applications in particle-physics contexts where fermion flavours are modelled as qutrits~\cite{Chernyshev:2024pqy,Thaler:2024anb}.

As a physical application, we identify the pion isospin triplet with a qutrit and use dispersively constrained elastic $\pi\pi$ phase shifts~\cite{Garcia-Martin:2011iqs,Beane:2021pion} to follow the non-local magic generated by the partial-wave $S$ matrix. This connects the qutrit formula directly to the broader programme of using entanglement and non-stabiliserness to probe the quantum-information structure of hadronic scattering~\cite{Beane:2018oxh,Low:2021ufv,Liu:2022grf,Bai:2023tey,Miller:2023ujx,Robin:2024oqc,Robin:2025ymq,Low:2026evp}.

The remainder of the paper is structured as follows. In Sec.~\ref{sec:qbits} we phrase the bipartite qubit case studied in Ref.~\cite{Qian:2025oit} in the language of invariants, preparing the reader for the extension to higher-dimensional qudits presented in Sec.~\ref{sec:general}. In Sec.~\ref{sec:analysis} we present closed-form formulae and analyse the $N=2,3,5$ cases. In Sec.~\ref{sec:pion-scattering} we apply the qutrit result to elastic pion scattering, before presenting our conclusions in Sec.~\ref{sec:conc}. Appendix~\ref{sec:num-schmidt-attainment} describes our numerical experiments validating the Schmidt-attainment hypothesis.

\section{Two-qubit non-local magic from local-unitary invariants}
\label{sec:qbits}

We begin with a pure two-qubit state
\begin{equation}
    \ket{\psi} = \sum_{j,k=0}^{1} x_{jk} \ket{jk} \ ;
\end{equation}
such a state always admits a Schmidt decomposition of the form
\begin{equation}
    \ket{\psi_\lambda} = \lambda_0 \ket{00} + \lambda_1 \ket{11} \ ,
\end{equation}
where $\lambda_i$ are the Schmidt coefficients, which can always be taken to be real and positive.

For a pure state the eigenvalues of the reduced density matrix encode the Schmidt coefficients. Indeed, from a practical point of view the Schmidt decomposition is obtained by finding the eigenvalues of the reduced density matrix of one of the subsystems. For a two-qubit system with subsystems $A$ and $B$, this is equivalent to
\begin{equation}
\label{eq:2qub-red-density-matrix}
\rho_X \sim \begin{bmatrix}
\lambda_0^2 & 0 \\
0 & \lambda_1^2
\end{bmatrix} \ ,
\end{equation}
where $\rho_{X}$ represents the reduced density matrix of subsystem $X$, for $X \in \{A,B\}$. Each normalised reduced density matrix satisfies
\begin{equation}
    \operatorname{Tr} \rho_{X} = 1
\end{equation}
and transforms as
\begin{equation}
    \rho_X \to U_X^\dagger \rho_X U_X
\end{equation}
under $U_A \otimes U_B$ transformations. Since the normalisation condition fixes the trace, there is a single invariant degree of freedom that can be taken to be the determinant of the reduced density matrix of Eq.~\eqref{eq:2qub-red-density-matrix}:
\begin{equation}
    e_2 \equiv \det \rho_{A} = \det \rho_{B} \ .
\end{equation}

Since the density matrix is Hermitian, it can be decomposed by projecting it over the set of matrices $\sigma_{\mu\nu} = \sigma_\mu \otimes \sigma_\nu$, where
\begin{equation}
\sigma_\mu \equiv \left\{
\mathbb{I}, \,
\sigma_x, \,
\sigma_y, \,
\sigma_z
\right\}
\end{equation}
and $\mu \in \{0,1,2,3\}$, which form a complete basis for $4\times4$ Hermitian matrices. Concretely, we take this projection to be
\begin{equation}
\rho = \frac{\sigma_{\mu\nu} \operatorname{Tr} \left[\sigma_{\mu\nu} \rho\right]}{\operatorname{Tr} \left[\sigma_{\mu\nu}\sigma_{\mu\nu}\right]}
    = \frac{1}{4}\sigma_{\mu\nu} \operatorname{Tr} \left[\sigma_{\mu\nu}\rho\right] \equiv \frac{1}{4}R_{\mu\nu}\sigma_{\mu\nu} \ ,
\end{equation}
where we understand an implied sum over repeated indices, and we have defined~\cite{Horodecki:1996qk,PhysRevA.52.4396}
\begin{equation}
    R_{\mu\nu} \equiv \operatorname{Tr} \left[\sigma_{\mu\nu}\rho\right] \ .
\end{equation}
The $R_{\mu\nu}$ are real as both $\rho$ and $\sigma_{\mu\nu}$ are Hermitian; these components transform in the same way as $\sigma_{\mu\nu} = \sigma_\mu \otimes \sigma_\nu$ under $U_A \otimes U_B$ transformations. As $\mathbb{I}$ is an $\mathrm{SU}(2)$ singlet and $\sigma_i$, with $i \in \{x,y,z\}$, transforms like a triplet, $R_{\mu\nu}$ can be decomposed as
\begin{equation}
\label{eq:Rblock}
R_{\mu\nu} =
\begin{bmatrix}
1 & a^{\mathsf{T}} \\
b & T
\end{bmatrix} \ ,
\end{equation}
where, in the sense of $\mathrm{SU}(2) \times \mathrm{SU}(2)$ representations, $a \in (\mathbf{1},\mathbf{3})$, $b \in (\mathbf{3},\mathbf{1})$, and $T \in (\mathbf{3},\mathbf{3})$, so they transform as $a\rightarrow O_B a$, $b \rightarrow O_A b$ and $T \rightarrow O_A T O_B^{\mathsf{T}}$, where the $O_{X}$ are elements of $\mathrm{SO}(3)$ that depend on $U_{X}$. Due to the fact that $\mathrm{SU}(2)$ is a double cover of $\mathrm{SO}(3)$, for any rotation $O_{X}$ there is always a unitary transformation $U_{X}$ of the qubit that gives such an $O_{X}$ for $R_{\mu\nu}$. This guarantees that the matrix $T$ can always be diagonalised with the right choices of $U_{A}$ and $U_B$.

Indeed, we highlight that the basis defined by the Schmidt decomposition diagonalises the matrix $T$. If we parametrise
\begin{equation}
    \lambda_0 = \cos\theta, \quad \lambda_1 = \sin\theta \ ,
\end{equation}
one gets
\begin{equation}
    e_2 = \frac{1}{4} \sin^2\left(2\theta\right),
\end{equation}
for the determinant of the reduced density matrices, and the subcomponents of $R_{\mu\nu}$ from Eq.~\eqref{eq:Rblock} take the form
\begin{equation}
    a = b = \begin{bmatrix}
    0 \\ 0 \\ \cos 2\theta
    \end{bmatrix}
\end{equation}
and
\begin{equation}
    T = \begin{bmatrix}
\sin 2 \theta & 0 & 0 \\
0 & -\sin 2 \theta & 0  \\
0 & 0 & 1
\end{bmatrix} \ .
\end{equation}

We proceed by adopting the second stabiliser R\'enyi entropy as our magic monotone~\cite{Leone:2021rzd}. In this case, the NLM can be expressed as
\begin{equation}
    \mathscr{M}_{AB} = \min_{U_A\otimes U_B}\left( -\ln \frac{\sum_{\mu\nu}\operatorname{Tr}\left[\sigma_{\mu\nu}\rho\right]^4}{\sum_{\mu\nu}\operatorname{Tr}\left[\sigma_{\mu\nu}\rho\right]^2}\right) \ ,
\end{equation}
which can be written in terms of the coefficients $R_{\mu\nu}$:
\begin{equation} \label{eq:qbit-non-local-magic}
    \mathscr{M}_{AB} = \min_{U_A\otimes U_B}\left( -\ln \frac{\sum_{\mu\nu} |R_{\mu\nu}|^4}{4}\right) \ ,
\end{equation}
where we have also specialised to the case of a pure state. The minimum over $U_A\otimes U_B$ occurs when
\begin{equation}
\label{eq:fourth-power-sum}
\sum_{\mu,\nu} R_{\mu\nu}^4
\end{equation}
is maximised,\footnote{Recall that $R_{\mu\nu} \in \mathbb{R}$.} and this maximum is attained in the Schmidt basis, where $T$ is diagonal~\cite{Qian:2025oit}.\footnote{A useful way to see this is to decompose the objective according to Eq.~\eqref{eq:Rblock}. The scalar component $R_{00}=1$ is invariant. The vectors $a_i=R_{0i}$ and $b_i=R_{i0}$ transform under local unitaries as $\mathrm{SO}(3)$ vectors, so their Euclidean norms are invariant. For fixed norm, the quartic sum $\sum_i a_i^4$ is maximised when only one component is non-zero, and likewise for $\sum_i b_i^4$; this is exactly what happens in the Schmidt basis. Similarly, the correlation-matrix elements $R_{ij}=T_{ij}$ can be written as $\sum_k (O_A)_{ik} l_k (O_B)_{jk}$ for suitable choices of $U_A$ and $U_B$, where $l_k$ are the singular values of $T$, which are invariants. The orthogonality constraints on $O_A$ and $O_B$ imply that mixing the singular-value weight across several entries cannot increase the sum $\sum_{ij} T_{ij}^4 = \sum_{ij}[\sum_k (O_A)_{ik} l_k (O_B)_{jk}]^4$, and the quartic sum is therefore maximised when $T$ is diagonal. Thus the maxima of all sectors are realised simultaneously in the Schmidt basis, yielding the global maximum of Eq.~\eqref{eq:fourth-power-sum}.}

For the Schmidt-form state $\ket{\psi}=\cos\theta\,\ket{00}+\sin\theta\,\ket{11}$, there are only six non-zero Pauli expansion coefficients $R_{\mu\nu}$: $R_{00}=R_{33}=1$, $R_{30}=R_{03}=\cos 2\theta$, and $R_{11}=-R_{22}=\sin 2\theta$. Evaluating the objective gives $\sum_{\mu\nu}R_{\mu\nu}^4=4-\sin^2(4\theta)$, which can be written in terms of the local-unitary invariant $e_2=\det\rho_A=\tfrac14\sin^2(2\theta)$ to give
\begin{equation} \label{eq:mn2}
    \mathscr{M} = -\ln\left(1-4e_2+16e_2^2\right)
\end{equation}
for the NLM of a two-qubit system.

\section{Extension to prime-dimensional qudits}
\label{sec:general}

In this section we extend the invariant-based construction of non-local magic from two qubits to bipartite systems of local dimension $N$, again specialising to the case of pure states. For $N=3,5$, we proceed under the Schmidt-attainment hypothesis: the global NLM minimum over local unitaries is attained by Schmidt-aligned states. We provide numerical evidence for this in Appendix~\ref{sec:num-schmidt-attainment}. Accordingly, the resulting closed-form expressions to be introduced in Sec.~\ref{sec:analysis} should be interpreted as conjectured formulae for prime $N$. Throughout this section we use $X \in \{A,B\}$ as a shorthand to label the subsystems $A$ and $B$.

We begin with a general pure state
\begin{equation}
    \ket{\psi} = \sum_{j,k=0}^{N-1} x_{jk} \ket{jk},
\end{equation}
which always admits a Schmidt decomposition
\begin{equation} \label{eq:schmidt}
    \ket{\psi_\lambda} = \sum_{i=0}^{N-1} \lambda_i \ket{ii} \ ,
\end{equation}
where the Schmidt coefficients $\lambda_i$ can be chosen real and non-negative.

As in Sec.~\ref{sec:qbits}, local unitary transformations $U_A \otimes U_B \in \mathrm{SU}(N)\otimes \mathrm{SU}(N)$ preserve the spectrum of the reduced density matrices, hence the $\rho_{X}$ have eigenvalues $\lambda_i^2$.\footnote{We note that the restriction to $\mathrm{SU}(N)$ here is made without loss of generality, since the omitted $\mathrm{U}(1)$ factors contribute only a global phase.} For $N>2$, the reduced spectrum contains $N-1$ independent invariants (after the normalisation constraint $\Tr \rho_{X} = 1$); our strategy is therefore to organise these invariants, rewrite the relevant fourth-order correlator in a generalised Pauli basis, and then identify the structure that controls the NLM minimisation.

Generalising the notation from the previous section, we define
\begin{align}
    p_n &\equiv \operatorname{Tr}\!\left(\rho_{X}^n\right) \ ,\\
    e_N &\equiv \det \rho_{X} \ ,
\end{align}
which are invariants under local unitary transformations. To express these in terms of Schmidt coefficients, we introduce symmetric and cyclic sums: for a list of exponents $a = (a_1,\dots,a_N)$, we have
\begin{align}
    s_{a_1\cdots a_N} &\equiv \sum_{\pi\in S_N}\prod_{j=1}^{N}\lambda_{\pi(j)}^{a_j} \ , \\
    c_{a_1\cdots a_N} &\equiv \sum_{\pi\in C_N}\prod_{j=1}^{N}\lambda_{\pi(j)}^{a_j} \ ,
\end{align}
where $S_N$ is the permutation group and $C_N$ its cyclic subgroup. For the symmetric sums $s$, the sum is understood to run over the distinct permutations of the exponents $a_i$, so that each distinct monomial is counted exactly once; for the cyclic sums $c$, all $N$ shifts are summed. Adopting notation that omits zero exponents when unambiguous, e.g.\ $s_{200}=s_2$, the corresponding relations are
\begin{align}
    p_n &= \sum_{i=0}^{N-1}\lambda_i^{2n} = s_{2n} = c_{2n} \ ,\\
    e_N &= \prod_{i=0}^{N-1}\lambda_i^2
    = s_{\!\!\!\underbrace{\scriptstyle 2\cdots 2}_{N\ \text{times}}}
    = \frac{1}{N} c_{\!\!\!\underbrace{\scriptstyle 2\cdots 2}_{N\ \text{times}}} \ .
\end{align}
Moreover,
\begin{equation}
    p_{1/2} = \operatorname{Tr}\!\left(\sqrt{\rho_{X}}\right)
    = \sum_{i=0}^{N-1}\lambda_i = s_1 = c_1 \ ,
\end{equation}
where $\sqrt{\rho_{X}}$ denotes the positive square root of $\rho_{X}$, which is well-defined because $\rho_{X}$ is Hermitian and positive semidefinite.

To connect with the NLM objective, we construct a qudit analogue of the qubit coefficients $R_{\mu\nu}$ introduced in Sec.~\ref{sec:qbits}. This requires the usual generalisations of the Pauli operators:
\begin{equation}
    X\ket{j} = \ket{j+1 \bmod N},
    \quad
    Z\ket{j} = \omega^j\ket{j},
\end{equation}
where $\omega=e^{2\pi i/N}$, so that $ZX=\omega XZ$. The two-qudit operator basis is then
\begin{equation}
    P_{abcd}\equiv (X^a Z^b)\otimes(X^c Z^d) \ ,
\end{equation}
with $a,b,c,d\in\{0,\dots,N-1\}$.\footnote{We omit the customary phase factor in the definition of $P_{abcd}$, since it cancels in $|\Tr(\rho P_{abcd})|^4$.} Expanding $\rho$ on this basis gives
\begin{equation}
    \rho=\frac{1}{N^2}\sum_{a,b,c,d} t_{abcd}P_{abcd} \ ,
\end{equation}
where
\begin{equation}
    t_{abcd} \equiv \operatorname{Tr}(\rho P_{abcd})
                = \sum_{j,k=0}^{N-1} x_{jk} x_{j+a,k+c}^* \omega^{b j + d k} \ .
\end{equation}
This allows us to write the NLM minimand as
\begin{equation}
  -\ln \left( \frac{1}{N^2} \sum_{a,b,c,d} |t_{abcd}|^4 \right) \ ,
  \label{eq:qudit-fourth-sum}
\end{equation}
in analogy with Eq.~\eqref{eq:qbit-non-local-magic}. The coefficients $t_{abcd}$ are then the qudit analogue of the qubit coefficients $R_{\mu\nu}$. The key difference is that, unlike $R_{\mu\nu}\in\mathbb{R}$, the coefficients $t_{abcd}$ are generally complex because $P_{abcd}$ is non-Hermitian for $N>2$.

\begin{hypothesis}
  \label{hyp:schmidt-attainment}
  Let $\ket{\psi}\in\mathbb{C}^{N}\otimes\mathbb{C}^{N}$ be a normalised pure state with Schmidt coefficients $\lambda=(\lambda_0,\ldots,\lambda_{N-1})$, and define the computational-basis Schmidt alignments
  \begin{equation}
    \mathcal{S}_{\lambda}
    =
    \left\{
      \ket{\psi_{\lambda,\pi}}
      =
      \sum_{j=0}^{N-1}\lambda_{\pi(j)}\ket{jj}
      :~\pi\in S_N
    \right\}.
  \end{equation}
  Schmidt attainment holds in dimension $N$ if, for every such state,
  \begin{equation}
    \min_{U_A,U_B\in\mathrm{SU}(N)}
    M\left[(U_A\otimes U_B)\ket{\psi}\right]
    =
    \min_{\ket{\phi}\in\mathcal{S}_{\lambda}}M(\ket{\phi}),
    \label{eq:schmidt-attainment}
  \end{equation}
  where $M$ is the second stabiliser R\'enyi entropy. Equivalently, Schmidt attainment holds if and only if the fourth-moment sum $\sum_{abcd}|t_{abcd}|^4$ attains its maximum over the local-unitary orbit of $\ket{\psi}$ at a state in $\mathcal{S}_{\lambda}$.
\end{hypothesis}

Schmidt attainment is exact for $N=2$. We find numerical support for it at $N=3,5$, but counterexamples at $N=4$; we therefore conjecture Eq.~\eqref{eq:schmidt-attainment} for prime local dimension. The apparent distinction between prime and composite dimension here echoes several structures familiar from phase-space formulations of qudit magic~\cite{Gross:2006wigner,Veitch:2012wigner,Delfosse:2017wigner}. 

Under the Schmidt-attainment hypothesis, we evaluate the correlator in the Schmidt-aligned form $x_{jk}=\lambda_j\delta_{j,k}$, which gives
\begin{equation}
    t_{abcd}
    = \sum_{j=0}^{N-1} \lambda_j \lambda_{j+a} \omega^{(b+d)j} \delta_{ac} \ .
\end{equation}
The Kronecker delta enforces $c=a$, so only those components contribute in Eq.~\eqref{eq:qudit-fourth-sum}. This also gives a block-diagonal structure for $t$, with $N$ blocks of dimension $N\times N$. In this case, the argument of the logarithm from Eq.~\eqref{eq:qudit-fourth-sum} is a useful quantity that we further define for use later:
\begin{equation} \label{eq:fn}
F_N(\lambda) \equiv \frac{1}{N^2} \sum_{a,b,c,d} |\Tr (\rho_\lambda P_{abcd})|^4 \ ,
\end{equation}
where $\rho_\lambda = | \psi_\lambda \rangle \langle \psi_\lambda |$. The fourth power of the modulus reads
\begin{equation}
  \begin{split}
    |t_{abad}|^4
    &= \sum_{j,k,p,q=0}^{N-1}
      \lambda_j \lambda_{j+a} \lambda_k \lambda_{k+a}
      \lambda_p \lambda_{p+a} \lambda_q \lambda_{q+a} \\
    &\quad \cdot \omega^{(b+d)(j-k+p-q)} \ .
  \end{split}
\end{equation}
In the remaining sum over $b$ and $d$, the phases are projected by
\begin{equation}
    \sum_{b,d=0}^{N-1}
    \omega^{(b+d)(j-k+p-q)}
    = N^2 \delta_{j-k+p-q,\,0\ (\mathrm{mod}\ N)} \ .
\end{equation}
After dividing by the normalisation factor $N^2$, only combinations satisfying
\begin{equation}
    j - k + p - q = 0 \pmod N
\end{equation}
survive.  This condition allows one to express $q$ as a function of the other indices:
\begin{equation}
    q = j - k + p \pmod N \ ,
\end{equation}
reducing the inner four-index sum to a triple sum over $(j,k,p)$, with an outer sum over $a$. The final expression, after performing the sums over $b$, $d$, and $c$, is then
\begin{equation} \label{eq:tab4}
    \sum_{a=0}^{N-1}
    \sum_{j,k,p=0}^{N-1}
    \lambda_j \lambda_{j+a}
    \lambda_k \lambda_{k+a}
    \lambda_p \lambda_{p+a}
    \lambda_{\,j-k+p} \lambda_{\,j-k+p+a} \ ,
\end{equation}
where all indices are taken modulo $N$. Each term corresponds to a cyclic combination of indices, and the remaining sum over $a$ ensures that only cyclically invariant patterns of the exponents contribute to the final expression.

For fixed Schmidt coefficients $\lambda=(\lambda_0,\ldots,\lambda_{N-1})$, the value of Eq.~\eqref{eq:tab4} is just $F_N(\lambda)$ from Eq.~\eqref{eq:fn}. Since local unitaries include permutations of the Schmidt basis, the Schmidt-attained NLM is obtained by evaluating $F_N$ over the inequivalent permutations of $\lambda$ (identifying orderings related by cyclic shifts) and taking the largest value:
\begin{equation}
    \mathscr{M}_{N} = -\ln\left(\max_{\lambda^\prime} F_N(\lambda^\prime)\right) \ ,
\end{equation}
where $\lambda^\prime$ runs over permutations of $\lambda$ modulo cyclic relabelling.

\section{Results for $N=2,3,5$ and numerical analysis}
\label{sec:analysis}

In the following we present simple formulae for two-qudit systems of low dimension and study the landscape of NLM produced in these cases, in order of increasing local dimension.

The closed-form expressions for the argument of the logarithm entering the NLM for $N=2,3,5$ are listed in Table~\ref{tab:closed-expr}. The analogous Schmidt-aligned expression for $N=4$, which does not reproduce the global local-unitary minimum in general, is given in Eq.~\eqref{eq:n4-schmidt-expression} of Appendix~\ref{sec:num-schmidt-attainment}.

For qubits, the corresponding formula reproduces Eq.~\eqref{eq:mn2} and the results of Ref.~\cite{Qian:2025oit}. A proof that the NLM minimum is attained in the Schmidt-aligned form in this case is found in the fact that the matrix $T$ of Eq.~\eqref{eq:Rblock} can always be diagonalised by appropriate choices of $U_A$ and $U_B$, owing to the relationship between $\mathrm{SU}(2)$ and $\mathrm{SO}(3)$. The expression depends only on $e_2\in[0,1/4]$ and therefore satisfies $3/4 \leq F_2(\lambda) \leq 1$, which implies
\begin{equation}
  0 \leq \mathscr{M}_{N=2} \leq \ln \frac{4}{3}  \ .
\end{equation}

For $N=3$ our formula reads
\begin{equation} \label{eq:nlm-qutrits}
F_3(\lambda) = \frac{1}{9}\sum_{a,b,c,d}|t_{abcd}|^4 = 1-2p_2+2p_2^2+4e_3 s_1^2 \ .
\end{equation}
Since $e_3 s_1^2\ge 0$, we have $F_3(\lambda) \ge f(p_2) \equiv 1-2p_2+2p_2^2$, whose minimum over the allowed range $p_2\in[1/3,1]$ occurs at $p_2=1/2$ and equals $f(1/2)=1/2$. This bound is saturated by a rank-two equal Schmidt spectrum, e.g.\ $\lambda=2^{-1/2}(1,1,0)$, for which $e_3=0$ and $p_2=1/2$, and thus $F_{3,\min}=1/2$. Conversely, $F_3(\lambda)=1$ (and hence $\mathscr{M}_{N=3}=0$) both for product states, where $p_2=1$ and $e_3=0$, and for the maximally entangled spectrum $\lambda_i=1/\sqrt{3}$, where $p_2=1/3$ and $e_3 s_1^2=1/9$. The non-local magic in two qutrits therefore satisfies
\begin{equation}
  0 \leq \mathscr{M}_{N=3} \leq \ln 2 \ .
\end{equation}

\begin{figure}[t]
\centering
\includegraphics[width=0.935\columnwidth]{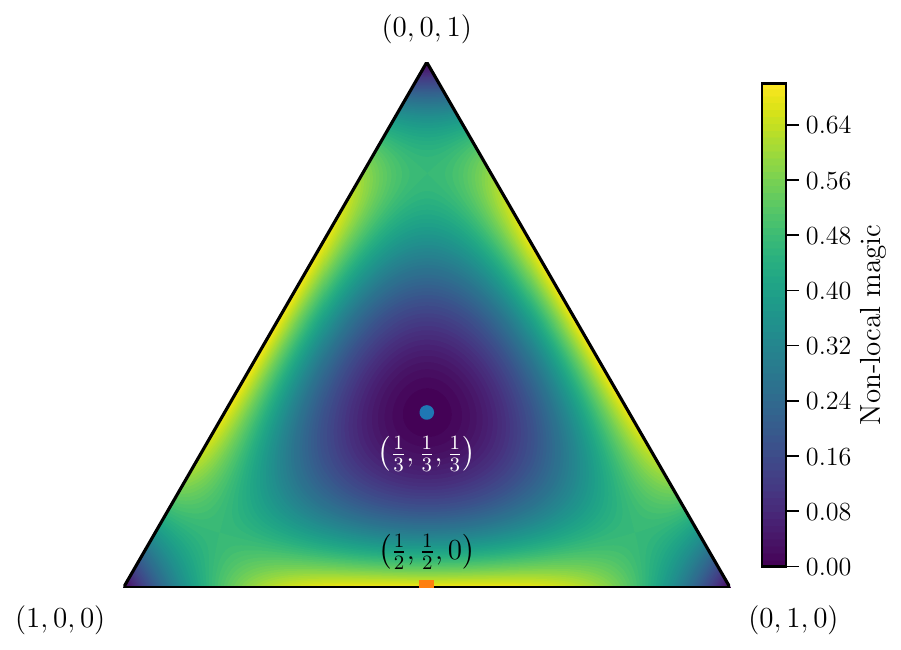}
\caption{Non-local magic for two qutrits across the canonical simplex of Schmidt probabilities $(\lambda_0^2,\lambda_1^2, \lambda^2_2)$, where $\sum_i \lambda_i^2=1$. The colour scale shows the non-local magic $\mathscr{M}_{N=3}=-\ln F_3(\lambda)$. The plot is symmetric under permutations of $(\lambda^2_0,\lambda^2_1,\lambda^2_2)$, appearing as a threefold symmetric pattern over the simplex. The minimum value $\mathscr{M}_{N=3}=0$ appears at product-state vertices and at the maximally entangled point $\lambda_i=1/\sqrt{3}$ (blue circle), while the maximum $\mathscr{M}_{N=3}=\ln 2 \approx 0.69 $ occurs at rank-two equal spectra, e.g.\ $\lambda = (1/\sqrt{2},1/\sqrt{2},0)$ (orange box), and permutations.}
\label{fig:qutrit-simplex}
\end{figure}

The dependence of the non-local magic on the Schmidt coefficients is visualised in Fig.~\ref{fig:qutrit-simplex} for the qutrit case. We use Schmidt probabilities $\lambda_i^2$ as canonical simplex coordinates, with $\sum_i \lambda_i^2=1$. We map these to Cartesian coordinates through
\begin{equation}
(x,y) = (\lambda^2_1+\tfrac{1}{2}\lambda_2^2, \tfrac{\sqrt{3}}{2} \lambda^2_2) \ ,
\end{equation}
so the vertices $(1,0,0)$, $(0,1,0)$, $(0,0,1)$ form an equilateral triangle. The landscape exhibits the expected overall $S_3$ permutation symmetry under relabelling of Schmidt components, and the aforementioned pattern of extrema in the non-local magic is clearly visible.

\begin{figure}[t]
\centering
\includegraphics[width=\columnwidth]{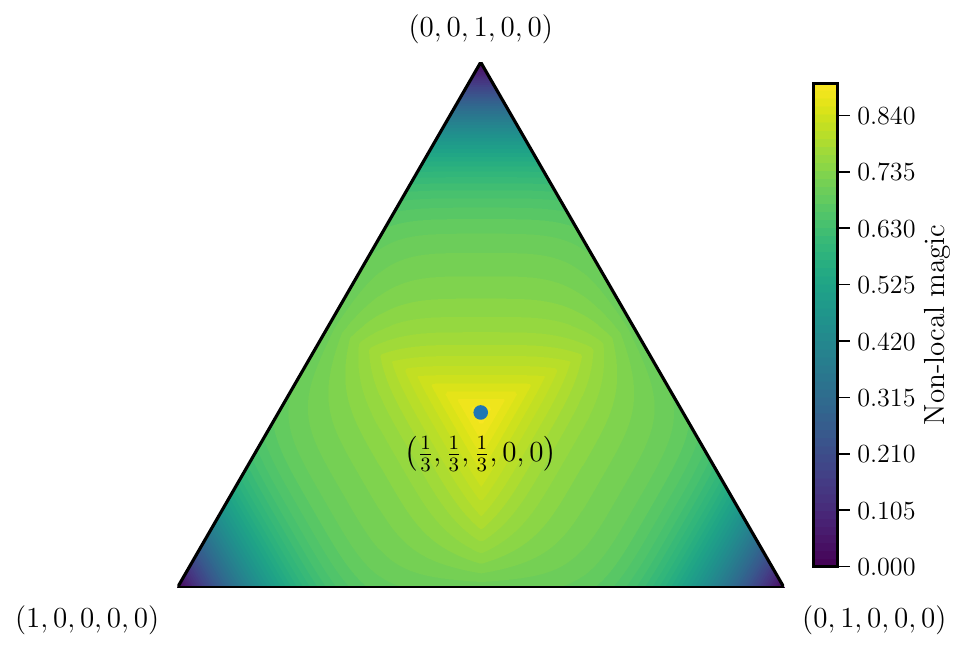}
\caption{Non-local magic for two ququints on the Schmidt-probability slice $(\lambda_0^2,\lambda_1^2,\lambda^2_2,0,0)$, where $\sum_i \lambda^2_i=1$. The colour scale shows the non-local magic $\mathscr{M}_{N=5}$. The plot is symmetric under permutations of $(\lambda^2_0,\lambda^2_1,\lambda^2_2)$, appearing as a threefold symmetric pattern over the triangular slice. The slice is chosen to feature the maximum value of $\mathscr{M}_{N=5}=\ln(27/11)\approx 0.90$, which occurs at $\lambda^2_0=\lambda^2_1=\lambda^2_2=1/3$ (blue circle).}
\label{fig:ququint-slice}
\end{figure}

For systems of two ququints, we likewise find that the minimum allowed non-local magic is $0$. This is attained for product states, for which one Schmidt coefficient equals $1$ and the others vanish, and also for the maximally entangled spectrum $\lambda_i=1/\sqrt{5}$, for which the Schmidt-based expression gives $F_5(\lambda)=1$. To identify the maximum allowed value of the non-local magic, we perform a numerical optimisation over Schmidt spectra. We find a value consistent with $\ln(27/11)$ to within numerical precision, associated with points along the boundary of Schmidt space corresponding to probabilities $(\lambda_0^2,\lambda^2_1,\lambda^2_2,\lambda^2_3,\lambda^2_4)=(1/3,1/3,1/3,0,0)$ and permutations. Thus,
\begin{equation}
  0 \leq \mathscr{M}_{N=5} \lesssim \ln\frac{27}{11} \ ,
\end{equation}
where the upper endpoint is supported numerically. We visualise a slice through this space for fixed $\lambda^2_3=\lambda^2_4=0$ in Fig.~\ref{fig:ququint-slice}. In this slice the central point $\lambda^2_0=\lambda^2_1=\lambda^2_2=1/3$ realises the maximal non-local magic.

We briefly comment on the relationship between the non-local magic and entanglement diagnostics. In our notation, the anti-flatness of the entanglement spectrum is
\begin{equation}
    \mathscr{F}(\ket{\psi}) = p_3-p_2^2 \ .
\end{equation}
For $n$ qubits, it is known that the \emph{local} magic (as measured by the stabiliser linear entropy~\cite{Leone:2021rzd}) is proportional to the anti-flatness averaged over the Clifford orbit~\cite{Tirrito:2023fnw}. Moreover, Ref.~\cite{Robin:2025ymq} recently reported an even more basic connection between the NLM and anti-flatness for two-qubit pure states: namely, $\mathscr{M}_{AB}^\text{lin}=4\mathscr{F}$, where $\mathscr{M}_{AB}=-\ln{(1-\mathscr{M}_{AB}^\text{lin})}$. This is readily proved in our formalism. By virtue of the normalisation constraint, we have $p_3=1-3e_2$ and $p_2=1-2e_2$, and hence $\mathscr{F}=e_2-4e_2^2$; referring to the $N=2$ row of Table~\ref{tab:closed-expr}, we see that this is exactly $\mathscr{M}_{AB}^\text{lin}/4$. This is consistent with the fact that, for two qubits, Eq.~\eqref{eq:mn2} can be written as a function of a single entanglement invariant, which can be taken to be the concurrence. For two qutrits, however, the situation changes: Eq.~\eqref{eq:nlm-qutrits} depends on more than the purity $p_2\equiv \Tr(\rho_X^2)$, and hence cannot be directly related to the concurrence alone~\cite{Iannotti2025entanglement}. We also verify numerically that $\mathscr{F}$ does not reproduce the value of $\mathscr{M}_{AB}^{\mathrm{lin}}/4$ implied by the $N=3$ formula. In fact, we find no linear relationship between $\mathscr{M}_{AB}^{\mathrm{lin}}$ and $\mathscr{F}$ in the qutrit case. Thus, while non-local magic is tightly linked to entanglement diagnostics for two qubits, this equivalence appears tied to the paucity of inequivalent invariants for $N=2$, and does not persist in general.

We emphasise a structural point relevant to interpreting these results. Since the NLM is defined by minimising over local unitaries, it is constant on each $U_A\otimes U_B$ orbit, and for bipartite pure states two states lie on the same orbit precisely when they share a Schmidt spectrum. It is therefore automatic, in any local dimension, that the NLM is some function of the entanglement spectrum; the non-trivial question is \emph{which} function. As we have seen, that function is not in general an entanglement monotone. 

\begin{table}[t]
\caption{Expressions entering $\exp(-\mathscr{M}_N)$ for two-qudit systems of local dimension $N$. The $N=2$ expression is exact, while the $N=3,5$ expressions rely on Hypothesis~\ref{hyp:schmidt-attainment} and are supported by the numerical tests in Appendix~\ref{sec:num-schmidt-attainment}.}
\label{tab:closed-expr}
\begin{ruledtabular}
\begin{tabular}{c l}
$N$ & $\exp(-\mathscr{M}_N)$ \\
\hline
$2$ & $1 - 4e_2 + 16e_2^2$ \\
$3$ & $1 - 2p_2 + 2p_2^2 + 4e_3 s_1^2$ \\
\multirow{2}{*}{$5$}
& $3 p_2^2 - 2 p_4 + 24 e_5^{1/2}s_{111} + 24 c_{2222} + 12 c_{242} + 12 c_{2204}$ \\
& ${}+ 8 c_{1331} + 8 c_{3113}$ \\
\end{tabular}
\end{ruledtabular}
\end{table}

\section{Pion scattering as a two-qutrit application}
\label{sec:pion-scattering}

As a physical application of Eq.~\eqref{eq:nlm-qutrits} we consider
elastic pion scattering $\pi\pi\to\pi\pi$ in the isospin-symmetric limit.  Each pion transforms in the
isospin-$1$ (adjoint) representation of the hadronic isospin group
$\mathrm{SU}(2)_I$.  We choose the basis
\begin{equation}
  \ket{1,\pm 1}=\ket{\pi^\pm},\qquad
  \ket{1,0}=\ket{\pi^0},
\end{equation}
where the first label denotes the single-pion isospin, while
the second is its third component, which here matches the electric charge.

The two qutrit
subsystems are identified with the two opposite-momentum modes in the
centre-of-mass frame, with product basis
$\ket{1,i;\mathbf{p}}\otimes\ket{1,j;-\mathbf{p}}$, where
$i,j\in\{+1,0,-1\}$. Addition of the two single-pion isospins gives $1 \otimes 1 = 0\oplus1\oplus2$. Isospin conservation makes the scattering operator diagonal in these two-pion isospin sectors.  In the elastic approximation, unitarity then implies that, for a fixed relative orbital angular momentum $\ell$~\cite{Beane:2021pion},
\begin{equation}
  S_\ell(s)=\sum_{I=0}^{2}e^{2i\delta_\ell^{(I)}(s)}P_I \ ,
  \label{eq:pion-partial-wave-s-matrix}
\end{equation}
where $P_I$ projects onto the sector of two-pion isospin $I$, $\delta_\ell^{(I)}$ is the corresponding phase shift, and $s$ is the squared centre-of-mass energy.  If $W$ swaps the two momentum modes, the projectors may be written as
\begin{equation}
  P_0 = \ket{0,0}\!\bra{0,0},~
  P_1 = \frac{\mathbf{1}-W}{2},~
  P_2 = \frac{\mathbf{1}+W}{2}-P_0 \ .
  \label{eq:pion-isospin-projectors}
\end{equation}
The $I=1$ subspace is antisymmetric under $W$, whereas the $I=0$ and $I=2$ subspaces are symmetric. Bose symmetry therefore requires  $\ell+I$ even. Setting the phases of the forbidden channels to zero gives~\cite{Beane:2021pion}
\begin{align}
\begin{aligned}
  S_0 &= e^{2i\delta_0^{(0)}}P_0+P_1+e^{2i\delta_0^{(2)}}P_2 \\
  S_1 &= P_0+e^{2i\delta_1^{(1)}}P_1+P_2 \ .
  \end{aligned}
  \label{eq:pion-s-and-p-matrices}
\end{align}
We evaluate these operators using the central values of the
dispersively constrained fits to data (CFD) of
Ref.~\cite{Garcia-Martin:2011iqs}, restricting the analysis to
$2M_\pi\leq\sqrt{s}\leq0.85~\mathrm{GeV}$.  This is the common domain
of the low-energy elastic parametrisations for the relevant $s$- and
$p$-wave phase shifts, which provide the dominant partial-wave
contributions in this energy range.

We begin by focusing on the fixed in-state $\ket{\pi^+\pi^-}$. The $p$-wave out-state has a simple form:
\begin{equation}
  S_1\ket{\pi^+\pi^-}
  =e^{i\delta_1^{(1)}}
  \left(
    \cos\delta_1^{(1)}\ket{\pi^+\pi^-}
    -i\sin\delta_1^{(1)}\ket{\pi^-\pi^+}
  \right) \ .
  \label{eq:pion-p-wave-output}
\end{equation}
Its Schmidt probabilities are
$(\cos^2\delta_1^{(1)},0,\sin^2\delta_1^{(1)})$.  Consequently,
$e_3=0$ and
\begin{equation}
  p_2
  =\cos^4\delta_1^{(1)}+\sin^4\delta_1^{(1)}
  =1-\frac{y}{2},
  \quad
  y\equiv\sin^2(2\delta_1^{(1)}) \ .
\end{equation}
Substitution into Eq.~\eqref{eq:nlm-qutrits} gives
$F_3=1-y+y^2/2$.  The NLM of the
$p$-wave output is therefore
\begin{equation}
  \mathscr{M}_{N=3}\!\left(S_1\ket{\pi^+\pi^-}\right)
  =-\ln\left(1-y+\frac{y^2}{2}\right) \ .
  \label{eq:pion-p-wave-nlm}
\end{equation}
This expression first reaches the maximum value $\ln 2$ at
$\sqrt{s}=0.710~\mathrm{GeV}$, where
$\delta_1^{(1)}=\pi/4$.  At the $\rho$-meson resonance
$\delta_1^{(1)}=\pi/2$, so $S_1=W$ and both entanglement and NLM vanish.

The $s$ wave, in contrast, also permits scattering into the neutral final state. Writing $a=e^{2i\delta_0^{(0)}}$ and $c=e^{2i\delta_0^{(2)}}$, we find
\begin{equation}
  S_0\ket{\pi^+\pi^-}
  =A\ket{\pi^+\pi^-}
   +B\ket{\pi^0\pi^0}
   +C\ket{\pi^-\pi^+} \ ,
  \label{eq:pion-s-wave-output}
\end{equation}
with
\begin{equation}
  A=\frac{2a+3+c}{6},\quad
  B=\frac{c-a}{3},\quad
  C=\frac{2a-3+c}{6} \ .
  \label{eq:pion-s-wave-amplitudes}
\end{equation}
Near $\sqrt{s}=0.596~\mathrm{GeV}$, the Schmidt probabilities are $(0.3303,0.3840,0.2857)$, for which the linear entropy is $1-\Tr\rho_A^2=0.6618$, close to its qutrit maximum $2/3$, whereas the NLM is only $\mathscr{M}_{N=3}=0.0145$.  Thus the physical $s$-wave trajectory realises a nearly maximally entangled state carrying almost no non-local magic.

\begin{figure*}[t]
  \centering
  \begin{minipage}[t]{0.49\textwidth}
    \centering
    \includegraphics[width=\linewidth]{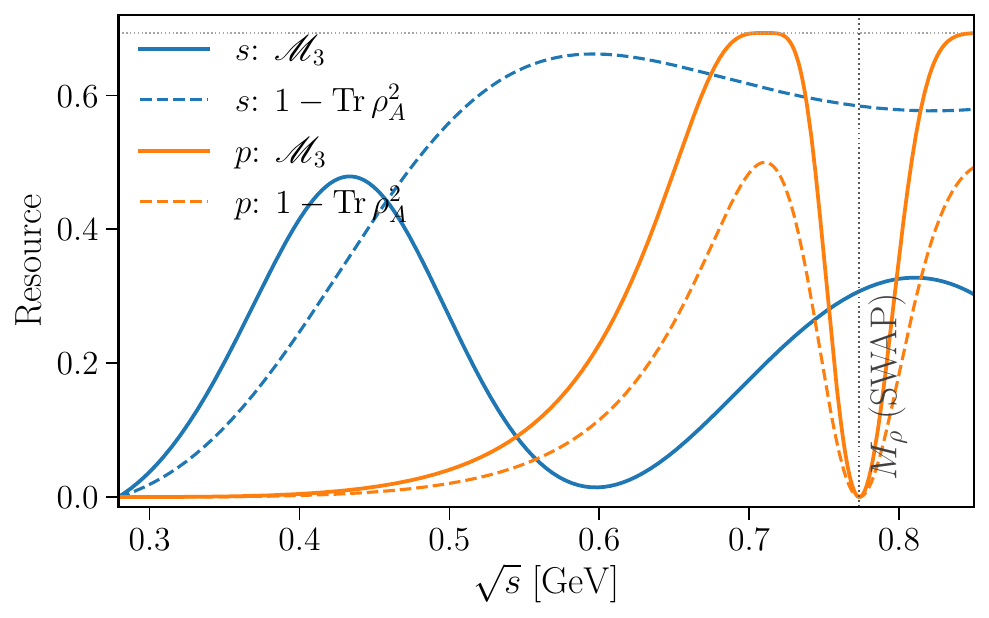}
  \end{minipage}
  \hfill
  \begin{minipage}[t]{0.49\textwidth}
    \centering
    \includegraphics[width=\linewidth]{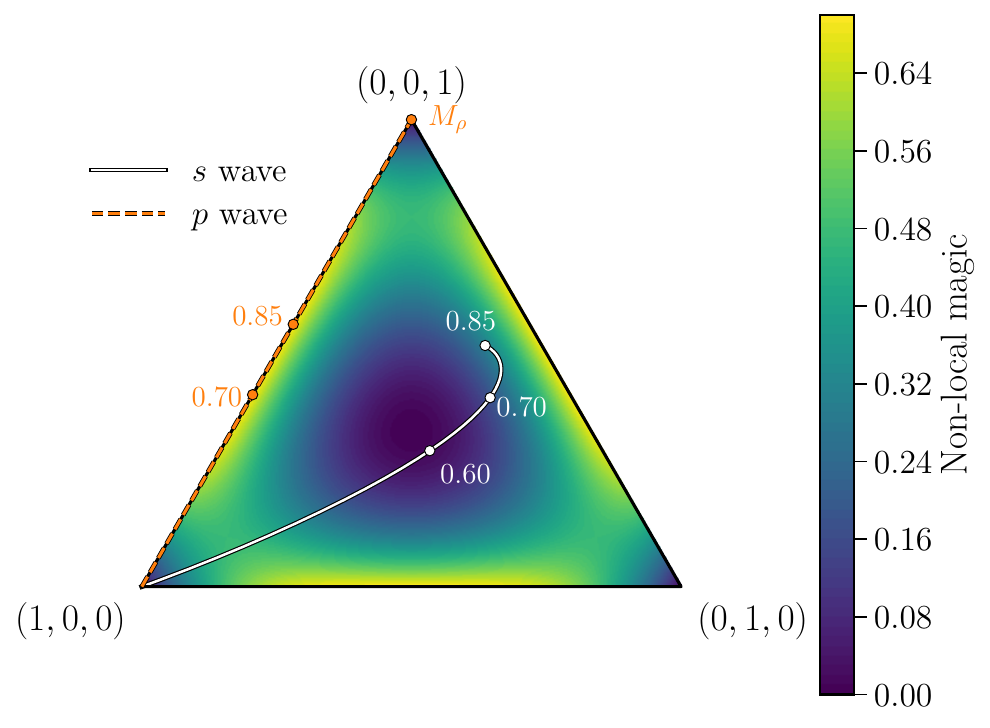}
  \end{minipage}
  \caption{Resources in the output state generated from the charged product input $\ket{\pi^+\pi^-}$ by elastic pion scattering.  Left: Two-qutrit NLM (solid) and linear entropy (dashed) for the $s$ (blue) and $p$ (orange) waves.  At the centre-of-mass energy $\sqrt{s}=M_\rho$, the $p$-wave scattering operator is SWAP, so both plotted resources vanish.  Near $\sqrt{s}=0.596~\mathrm{GeV}$, the linear entropy is close to its qutrit maximum of $2/3$, whereas the NLM is small. The dotted grey horizontal line shows $\ln 2$, the maximum NLM of a two-qutrit system, reached by the $p$ wave either side of the $\rho$ resonance.  Right: The corresponding trajectories on the simplex of Schmidt probabilities $(\lambda_{+-}^2,\lambda_{00}^2,\lambda_{-+}^2)$.  The $p$ wave (dashed, orange) is confined to the rank-two boundary, while the $s$ wave (solid, white) passes close to the uniform spectrum $(1/3,1/3,1/3)$, which is maximally entangled and has vanishing NLM.  Numerical trajectory labels give $\sqrt{s}$ in GeV. At $\sqrt{s}=M_\rho$, the $p$-wave trajectory reaches the product-state vertex $(0,0,1)$ and returns along the boundary towards the $(1,0,0)$ vertex.}
  \label{fig:pion-charged-resources}
\end{figure*}

All of these features are visualised in Fig.~\ref{fig:pion-charged-resources}, which displays the charged-input result in two complementary ways.  In the left plot we show the non-local magic and linear entanglement entropy for both $s$- and $p$-wave scattering. The $s$-wave NLM first reaches its maximum $0.4790$ at $\sqrt{s}=0.433~\mathrm{GeV}$ and then falls almost to zero near $0.596\,\mathrm{GeV}$, where the linear entropy is close to its maximum.  The $p$ wave instead reaches the absolute qutrit NLM bound $\ln 2$ at $0.710~\mathrm{GeV}$, returns to zero at $M_\rho$, where $S_1$ is the SWAP gate, and rises towards its second maximum above the resonance. The right plot explains this contrast geometrically, showing the trajectories within the Schmidt-probability simplex with increasing $\sqrt{s}$. The $p$ wave moves along a rank-two boundary and turns around at the $(0,0,1)$ vertex, whereas the $s$ wave enters the simplex interior and passes close to its central zero point, corresponding to maximal entanglement.

To test that these features are not peculiar to a single charge state, we also define the NLM-generating power over a finite stabiliser ensemble $\mathcal{E}$,
\begin{equation}
  \mathcal{P}_{\mathrm{NLM}}(S_\ell;\mathcal{E})
  =\frac{1}{|\mathcal{E}|}
   \sum_{\ket{\phi}\in\mathcal{E}}
   \mathscr{M}_{N=3}\!\left(S_\ell\ket{\phi}\right) \ .
  \label{eq:pion-nlm-power}
\end{equation}
We compare results obtained using either the 144 products of the 12 single-qutrit stabiliser states or the complete set of 360 pure two-qutrit stabiliser states~\cite{Robin:2025ymq}.  The product ensemble is the direct stabiliser analogue of the product-state average used for entanglement power. We have checked that our setup reproduces the analytic $p$-wave entanglement power $\frac14\sin^2(2\delta_1^{(1)})$ found in Ref.~\cite{Beane:2021pion}.

\begin{figure*}[t]
  \centering
  \includegraphics[width=0.70\textwidth]{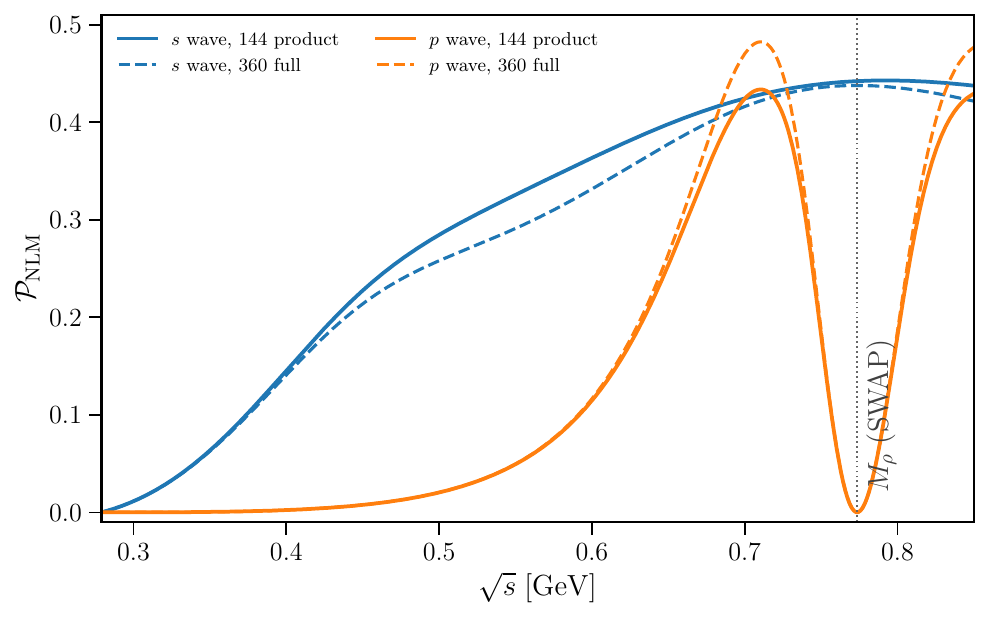}
  \caption{The figure shows the NLM-generating power of the $s$- and $p$-wave $\pi\pi\to\pi\pi$ $S$ matrices obtained from the CFD phase shifts. The $s$ and $p$ wave results are respectively shown in blue and orange. The mean is taken over either the 144 product stabiliser states (solid) or the set of all 360 pure two-qutrit stabiliser states (dashed). The $p$-wave power peaks before the $\rho$ resonance and vanishes exactly at $M_\rho$, where the $p$-wave scattering operator is the Clifford SWAP gate.}
  \label{fig:pion-nlm-power}
\end{figure*}

Fig.~\ref{fig:pion-nlm-power} shows the NLM-generating power of the
partial-wave $\pi\pi\to\pi\pi$ $S$ matrices obtained from the CFD phase
shifts.  We show averages over both the product and full stabiliser
ensembles and find the resulting curves to be closely aligned.  The
$s$-wave generating power remains broadly nonzero, while the $p$-wave
power peaks near $0.710~\mathrm{GeV}$ and vanishes exactly at the
$\rho$-meson resonance, where the $p$-wave operator is SWAP.  The product-
and full-ensemble peak values are respectively $0.4336$ and $0.4825$
for the $p$ wave, while the corresponding $s$-wave maxima are $0.4429$
and $0.4378$.

\section{Conclusions}
\label{sec:conc}

We have developed an invariant-based construction of non-local magic for bipartite pure states of qudits and obtained compact analytic expressions for prime local dimensions $N=3,5$. Our results depend on the Schmidt-attainment hypothesis: that the minimum over local unitaries defining the non-local magic is attained in the Schmidt-aligned family of states. Numerical comparisons between the Schmidt-based formulae and direct optimisation over $U_A \otimes U_B$ show strong agreement for $N=3,5$. In contrast, for $N=4$ the analogous Schmidt-based expression does not reproduce the global minimum across parameter space, although it often remains close. We find that for systems of two qutrits, the maximum allowed non-local magic is $\ln 2$, while for ququints numerical scans indicate a maximal value consistent with $\ln(27/11)$. We also found that relations between the non-local magic and other diagnostics such as the concurrence and the anti-flatness of the entanglement spectrum, which were present for systems of two qubits, generally do not extend past $N=2$. Overall, these results provide practical and largely analytic access to non-local magic in prime-dimensional two-qudit systems, and motivate a broader conjecture that Schmidt attainment persists for higher prime $N$.

Elastic pion scattering provides a concrete few-body illustration of the qutrit result. Scattering from an initial $|\pi^+ \pi^- \rangle$ state, the $p$ wave spans the full rank-two NLM range; it becomes SWAP at the $\rho$-meson resonance and reaches the maximum value of $\ln 2$ either side of this. The $s$ wave produces a nearly maximally entangled state for which the non-local magic is small.

There are several natural directions for future work. On the formal side, it would be valuable to establish the Schmidt-attainment hypothesis analytically, or else to characterise precisely the circumstances under which it fails. Our results also motivate the search for general invariant-based expressions in higher prime dimensions, together with refined approximations for composite dimensions where the Schmidt-aligned construction is not globally exact. Beyond bipartite pure states, it would be interesting to extend these ideas to mixed states, multipartite systems, bipartite unitary gates and many-body settings. Clarifying the relation between non-local magic and entanglement diagnostics in systems of higher local dimension is also an interesting research direction.

\begin{acknowledgments}
We thank Gary Mooney and C\'{e}dric Simenel for useful discussions. All authors are supported by the Australian Research Council grant CE200100008. MJW is further supported by the Australian Research Council grant DP220100007. ENVW acknowledges the support he has received for this research through the provision of an Australian Government Research Training Program Scholarship.
\end{acknowledgments}

\appendix

\section{Comparison with numerical optimisation over local unitaries}
\label{sec:num-schmidt-attainment}

\begin{figure*}[t]
\centering
\includegraphics[width=\textwidth]{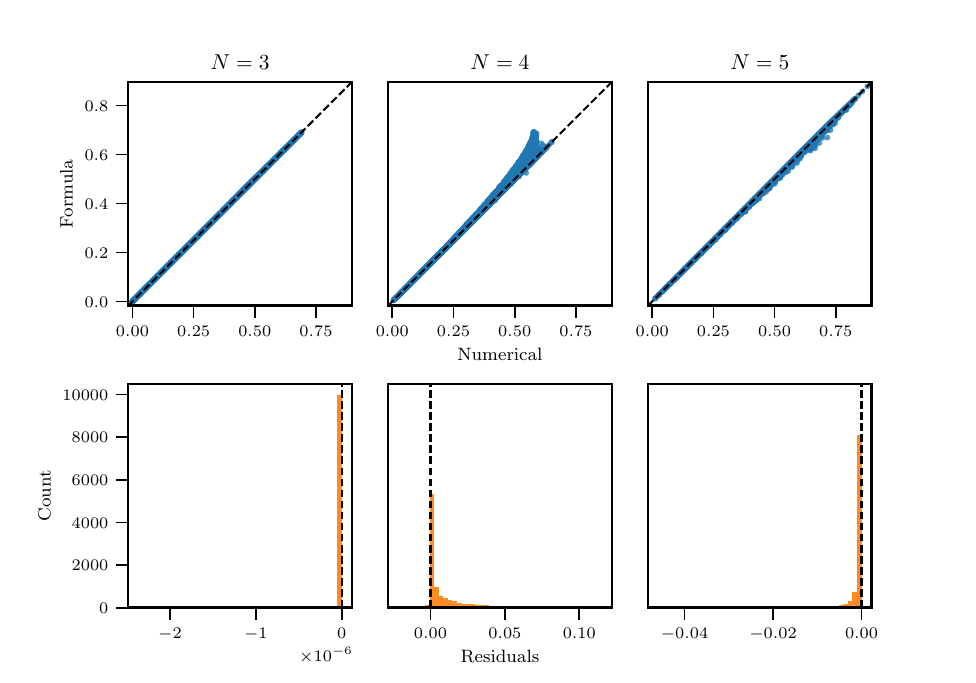}
\caption{Comparison of the non-local magic from direct numerical minimisation over $U_A \otimes U_B$ and from the Schmidt-based formula for $N=3,4,5$. For each $N$, we sample $10^4$ points in Schmidt-parameter space and perform multi-start L-BFGS optimisation with $n_{\mathrm{starts}}=50$ ($N=3$), $200$ ($N=4$), and $500$ ($N=5$), using $\texttt{maxiter}=300$ per start. Top row: formula value versus numerically minimised value, with the dashed line showing equality. Bottom row: residual histograms for $\mathscr{M}_{\mathrm{formula}}-\mathscr{M}_{\mathrm{numerical}}$. Negative residuals are consistent with incomplete convergence of the numerical minimisation; positive residuals indicate formula values above the numerical minimum. For $N=3,5$, no positive residuals are observed in the sampled points, supporting Schmidt attainment in prime dimensions. For $N=4$, positive residuals are present, showing that the analogous Schmidt-based expression does not reproduce the global minimum in general; nevertheless, it often lies close to the minimum.}
\label{fig:formula-vs-numerical-panels}
\end{figure*}

Below we present numerical evidence to motivate what we have dubbed the Schmidt-attainment hypothesis: that the minimum over local unitaries defining the NLM is attained in the Schmidt-aligned family of states for $N=3,5$. This motivates our conjecture that Schmidt attainment holds for pairs of qudits of prime local dimension more generally.

For comparison with the prime-dimensional cases, evaluating the same fourth moment on the Schmidt-aligned ansatz at $N=4$ gives
\begin{equation}
  F_4(\lambda)
  =
  3p_2^2-2p_4+120e_4
  +4c_{404}+12c_{242}+8c_{1331}.
  \label{eq:n4-schmidt-expression}
\end{equation}
We use $-\ln F_4$ only as a diagnostic Schmidt-aligned prediction; it is not asserted to equal the NLM. Our composite-dimensional numerical test is restricted to $N=4$.

To test this hypothesis, we sample $10^4$ sets of Schmidt coefficients by drawing spectra $\lambda\in\mathbb{R}_{\ge 0}^N$ uniformly on the positive orthant of the unit $(N-1)$-sphere and calculate (i) the value obtained by direct numerical minimisation over $U_A\otimes U_B$, and (ii) the corresponding Schmidt-aligned prediction, using Table~\ref{tab:closed-expr} for $N=3,5$ and Eq.~\eqref{eq:n4-schmidt-expression} for $N=4$.

The numerical minimisation is performed using the L-BFGS algorithm through Optax~\cite{deepmind2020jax} with $n_{\mathrm{starts}}=50$ ($N=3$), $200$ ($N=4$), and $500$ ($N=5$), using $\texttt{maxiter}=300$ per start.

Fig.~\ref{fig:formula-vs-numerical-panels} shows this comparison for $N=3,4,5$: the top row plots the result from our formula versus the numerically optimised values, while the bottom row histograms the residuals $\mathscr{M}_{\mathrm{formula}}-\mathscr{M}_{\mathrm{numerical}}$. For $N=3,5$ almost all residuals are numerically indistinguishable from $0$, with no positive residuals observed. Negative values are consistent with incomplete convergence of the numerical minimisation, which is more likely for the ququint scan for which the optimisation occurs over a much higher dimensional space.

For $N=4$, by contrast, we find that the analogous Schmidt-based expression does not reproduce the global minimum throughout the full parameter space, although the prediction from our formula is usually a good estimate. Since the Schmidt-aligned states lie on the local-unitary orbit of $\ket{\psi}$, the Schmidt-aligned value is a rigorous upper bound on the NLM for any $N$, and each positive residual therefore confirms that the Schmidt-aligned expression exceeds the true NLM by at least that amount, independently of the convergence of the numerical minimisation. The residuals $\Delta\mathscr{M} \equiv \mathscr{M}_{\mathrm{formula}}-\mathscr{M}_{\mathrm{numerical}}$ have mean $0.010$, standard deviation $0.017$, and maximum $0.115$; approximately $72\%$ of the sampled spectra satisfy $|\Delta\mathscr{M}| < 0.01$.

\begin{figure*}[t]
\centering
\includegraphics[width=0.8\textwidth]{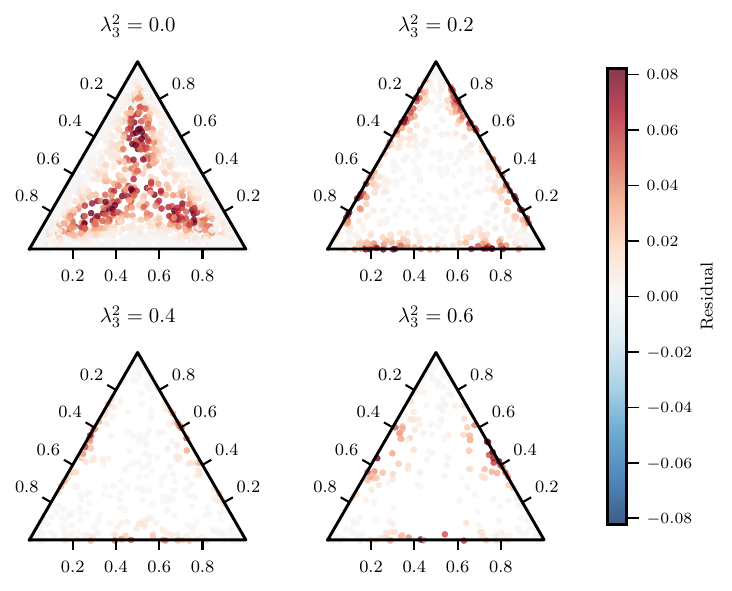}
\caption{Residual map for the $N=4$ Schmidt-based formula across projected slices of Schmidt-probability space. Shown is the residual $\mathscr{M}_{\mathrm{formula}}-\mathscr{M}_{\mathrm{numerical}}$ for fixed $\lambda_3^2 \in \{0,0.2,0.4,0.6\}$. For each panel, points are selected in a band $| \lambda_3^2-(\lambda_3^2)_*|\le 0.02$; the remaining components $(\lambda_0^2,\lambda_1^2,\lambda_2^2)$ are then normalised by $1-\lambda_3^2$, and the data are plotted on the canonical simplex. The largest residuals occur where one Schmidt component is small while the remaining components are nearly equal, i.e.\ near the rank-3 maximally entangled face $\lambda^2 \propto (0,1,1,1)$ and its permutations; near-product spectra remain well described. Very few points have negative residuals, and all such points are localised to the vertices, which represent $(\lambda_0^2,\lambda_1^2,\lambda_2^2)=(1,0,0)$ and permutations.}
\label{fig:n4-projections}
\end{figure*}

In order to better map the regions in $\lambda^2$-space in which the $N=4$ formula is trustworthy, we plot the residual $\Delta\mathscr{M}$ on slices across fixed $\lambda_3^2\in\{0,0.2,0.4,0.6\}$ in Fig.~\ref{fig:n4-projections}. For each value we select points in a narrow band $|\lambda_3^2-(\lambda_3^{2})_{*}|\leq 0.02$, normalise $(\lambda_0^2,\lambda_1^2,\lambda_2^2)$ by $1-\lambda_3^2$, and display the points on the canonical simplex. The largest residuals concentrate where one Schmidt probability is small while the remaining three are nearly equal, i.e.\ near the rank-3 maximally entangled face $\lambda^2\propto(0,1,1,1)$ and its permutations, visible as the large residuals filling the interior of the $\lambda_3^2=0$ slice; near-product spectra remain well described.

We comment briefly on the character of this breakdown at $N=4$. Interestingly, the Schmidt-aligned states remain locally optimal even where they fail globally. For spectra sampled both from the failure region and at random, we verify numerically that the NLM minimand of Eq.~\eqref{eq:qudit-fourth-sum} is stationary at the optimal Schmidt alignment, with positive-semidefinite Hessian under all local-unitary perturbations. Moreover, on sampled spectra with $\Delta\mathscr{M}>0.05$ (characterising the failure region), the numerical values cluster around $0.56$ and never exceed $\approx 0.59$, while the Schmidt-aligned prediction continues to grow. From this we conclude that the Schmidt-aligned states occupy the bottom of their own basin of the optimisation landscape, while the true optimum lies in a different basin, on a distinct family of stationary points. A detailed characterisation of this family is deferred to future work.

\bibliographystyle{apsrev4-2}
\bibliography{main}

\end{document}